\documentclass{iau}
\usepackage{graphicx}

\title{Next Generation Deep 2$\mu$ Survey}
\author{Jeremy Mould}
\affiliation{$^1$Centre for Astrophysics and Supercomputing, Swinburne University, Hawthorn 3122, Australia\\
[\affilskip]
$^2$ARC Centre of Excellence for All-sky Astrophysics (CAASTRO)\\ email: {\tt jmould@swin.edu.au}\\

}

\pubyear{2012}
\volume{288}  
\pagerange{119--126}
\jname{Astrophysics from Antarctica}
\editors{M.G. Burton, X. Cui \& N.F.H. Tothill, eds.}
\begin{document}

\maketitle

\begin{abstract}

There is a major opportunity 
 for the KDUST 2.5m telescope to carry out the next generation IR survey.
A resolution of 0.2 arcsec is obtainable from Dome A
over a wide field. This opens a unique discovery space during the 2015-2025 decade.

A next generation 2$\mu$ survey will feed JWST with serendipitous targets for spectroscopy, 
including spectra and images of the first galaxies.

\end{abstract}

\firstsection 
\section{Introduction: the state of the art of infrared surveys}
UKIDSS\footnote{www.ukidss.org}
has surveyed 7500 square degrees of the Northern sky, extending over both high and low Galactic latitudes, in the JHK bandpasses to K=18.3. 
This is three magnitudes deeper than 2MASS. 
UKIDSS has provided a near-infrared SDSS
and a panoramic atlas of the Galactic plane.  
UKIDSS is actually five surveys, including
two deep extragalactic elements, one covering 35 sq deg to K = 21, and the other reaching K = 23 over 0.77 sq deg.

VIKING-VISTA\footnote{www.astro-wise/projects/VIKING} is a  kilo-degree infrared galaxy survey. 
The VIKING survey will image 1500 sq deg in Z, Y, J, H, and Ks to a limiting magnitude 1.4 mag beyond the UKIDSS Large Area Survey. It will furnish 
very accurate photometric redshifts, especially at z $>$ 1, an important step in weak lensing analysis and observation of baryon acoustic oscillations.
Other science drivers include the hunt for high redshift quasars, galaxy clusters, and the study of
           galaxy stellar masses.

Mould (2011) offers a summary of the prospects for improving on these surveys using the KDUST 2.5m telescope.

\section{KDUST camera architecture}
The simplest option for a focal plane array is a Teledyne HgCdTe 2048$^2$.
A better option is 4096$^2$ or 2 x 2 (8.5 arcmin field).
ANU has delivered two such cameras to the Gemini Observatory (McGregor \etal~ 2004, McGregor \etal~ 1999).
The KDUST focal plane scale is appropriate without change.
JHK and Kdark filters would be required.

Plan B is for a Sofradir SATURN SW HgCdTe SWIR.
However, these detectors have 150 electrons read noise and would require long exposures to overcome readout noise. 
Nevertheless they are feasible Plan B detectors for broadband survey work. Mosaicing many detectors is also acceptable for survey work, and, after mosaicing the focal plane, plan A and plan B detectors are fairly similar in cost.

\section{The Antarctic advantage}
Above the ground layer turbulence one obtains almost diffraction limited images
over a wide field with low 2$\mu$ background.
This combination is only available from
the Antarctic plateau, high altitude balloons and space.
The competition, then, is space. We confine ourselves to WFIRST, since the ESA Euclid mission observes at H band, but not at K.
\vspace*{0.5 truecm}

\leftline{\bf Advantages of WFIRST} 
\begin{itemize}
\item Top ranked in ASTRO 2010 (Blandford 2009)
\item Broader band possible, e.g. 1.6-3.6$\mu$.
\item No clouds
\end{itemize}
\leftline{\bf Disadvantages of WFIRST}
\begin{itemize}
\item 3 year mission lifetime
\item Earliest launch 2025
\item Order of magnitude higher cost
\end{itemize}

Provided the US NRO supplies a 2.5m mirror, the following Astro2010-era disadvantages are no longer in effect.
\begin{itemize}
\item Smaller aperture, 1.5 metre
\item Lower resolution
\item 200 nJy limit vs 70 nJy with KDUST
\end{itemize}

\section{Science case}
An excellent science case for a 2.5m Antarctic telescope is presented by Burton \etal~ (2007). 
A further science case is that of WFIRST (Green \etal~ 2012)
\begin{itemize} 
\item Kuiper Belt census and properties
\item Cluster and Star-Forming-Region IMFs to planetary mass
\item The H$_2$ kink in star cluster CMDs 
\item The most distant Star-Forming-Regions in the Milky Way
\item Quasars as a Reference Frame for Proper Motion 	Studies
\item Proper Motions and parallaxes of disk and bulge Stars
\item Cool white dwarfs as Galactic chronometers
\item Planetary transits
\item Evolution of massive Galaxies: formation of red sequence galaxies
\item Finding and weighing distant, high mass clusters of galaxies
\item Obscured quasars
\item Strongly lensed quasars
\item High-redshift quasars and Reionization
\item Faint end of the quasar luminosity function
\item Probing Epoch of Reionization with Lyman­$\alpha$ emitters
\item Shapes of galaxy haloes from gravitational flexion 
\end{itemize}

\vspace*{5mm}
To focus on one of these areas, it is interesting to note the discovery space in the investigation of the epoch of reionization:

\begin{itemize}
\item 1$\mu$ band dropouts at z = 1.1/0.09 -1 = 11
\item J band dropouts at z = 1.4/0.09 -1 = 14
\item Galaxies with 10$^8$ year old stellar pops at z = 6
\item Pair production SNe (massive stars) at M$_K $= -23
\item Activity from the progenitors of supermassive black holes
\item Dark stars, see Ilie \etal~ (2012)
\item Young globular clusters with 10$^6$ year free fall times and M/L approaching 10$^{-4}$
\item Rare bright objects requiring wide field survey, then JWST, TMT, EELT or GMT spectra.
\end{itemize}

\section{The next steps}
The first question is whether this project is compatible with KDUST 2.5 (Cui 2010, Zhao \etal~ 2011). 
Assuming it is, we need to 
finalize the IR camera configuration, find IR camera partners, such as U. Tasmania,
Swinburne University, UNSW, AAO/Macquarie University, Texas A\&M, ANU and University of Melbourne.
We then need to flowdown the science to camera requirements.

To maximize advantage over VISTA, the speed of a survey to a given magnitude (inverse of the number of years to complete 1 sr) is a factor of $\sim$9. 
The goal is to increase this and get a full order of magnitude (or better).
Perhaps we should move from K to Kdark, when we have accurate measurements of the relevant backgrounds.
We could consider adding a reimager to KDUST and undersample a bit. Alternatively a slightly faster secondary on KDUST could be entertained.
For Sofradir chips the minimum exposure time is larger to overcome readout noise. For a background of 0.1 mJy/sq arcsec, the photon rate is half a photon per sec. This requires $>$2000 sec exposures for photon noise to double the Sofradir readout noise. (70\% QE assumed.)

A construction and operations schedule tentatively would be:
\begin{itemize}
\item January 2015 ARC LIEF funding, followed by Preliminary Design Review
\item 2016 Texas A \& M purchases Teledyne arrays; ANU purchases dewar and filters
\item 2016 Integrate and test focal plane at ANU or AAO
\item January 2017 Integrate telescope/ camera in Fremantle
\item 2018-2021 operations (within the international antarctic science region) at Kunlun Station
\item 2022 return of focal plane to the USA.
\end{itemize}

\vspace*{5mm}
This schedule is set by the time to manufacture and test the KDUST telescope in China.
If it slipped a year or two, so could the instrument schedule, although we do have
the precedent of GSAOI, where the camera was ready years before the adaptive optics.
The Centre for All-Sky Astrophysics in Australia and the proposed Joint Australia-China research centre would provide a very appropriate context for this collaboration. 
At Dome A astronomy can have another world class astrophysics enterprise in Antarctica yielding major results.
													 
\acknowledgements
I would like to thank our Chinese colleagues for hosting a workshop at the Institute of High Energy Physics
of the Chinese Academy of Sciences in Beijing in November 2011,
where a number of these ideas were developed.
Survey astronomy 
is supported by the Australian Research Council through CAASTRO\footnote{www.caastro.org}.
The Centre for All-sky Astrophysics is an Australian Research Council Centre of Excellence, funded by grant CE11001020.

\section*{References}
\noindent Blandford, R. 2009, AAS, 213, 21301\\
Burton, M. \etal~ 2007, PASA 22, 199. \\
Cui, Xiangqun	2010, Highlights of Astronomy, 15, 639\\
Green, J. \etal~ 2012, astro-ph 1208.4012\\
Ilie, C., Freese, K., Valluri, M., Iliev, I., \& Shapiro, P. 2012, MNRAS, 422, 2164\\	
McGregor, P., Hart, J., Stevanovic, D., Bloxham, G., Jones, D., Van Harmelen, J., Griesbach, J., Dawson, M., Young, P., \& Jarnyk, M.
2004, SPIE, 5492, 1033\\
McGregor, P. J., Conroy, P., Bloxham, G., \& van Harmelen, J. 1999 PASA, 16, 273\\
Mould, J. 2011, PASA, 28, 266\\
Thompson, R. \etal~ 2007, ApJ, 657, 669\\
Zhao, Gong-Bo, Zhan, Hu, Wang, Lifan, Fan, Zuhui, Zhang, Xinmin 2011, PASP, 123, 725\\

\begin{figure}[b]
\begin{center}
 \includegraphics[width=4in]{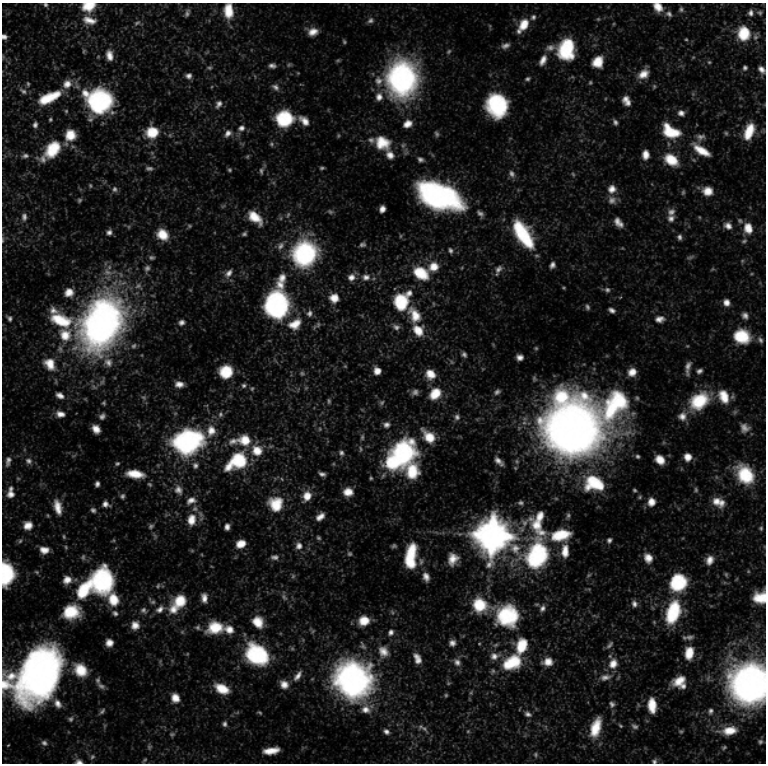} 
 \caption{The proposed survey will reach to within a magnitude of the NICMOS deep field (Thompson \etal~ 2007)
with similar resolution, but cover many steradians. The field shown here is a few arc minutes. {\it
Image courtesy NASA, Hubble Space Telescope, University of Arizona}. }
\end{center}
\end{figure}



\begin{discussion}

\discuss{Charling Tao}{There are problems with the persistence of Sofradir arrays.}
\discuss{Jeremy Mould}{There are strategies for dealing with the persistence.
But, thank you, this will need to be investigated for the Plan B detectors.}
\discuss{Hans Zinnecker}{What about L and M band?}
\discuss{Jeremy Mould}{This is feasible. However, even in the Antarctic the thermal background
comes roaring up and one becomes uncompetitive with space for broadband.}

\end{discussion}
\end{document}